\documentclass[manuscript]{aastex}

\shorttitle{Phasing telescopes}
\shortauthors{Takefuji et al.}

\usepackage{natbib}
\usepackage{amsmath} 
\usepackage{lscape} 
\usepackage{url}
\begin{document}

\title{ Development of the Phase-up Technology of the Radio Telescopes: 6.7 GHz Methanol Maser Observations with Phased Hitachi 32 m and Takahagi 32 m Radio Telescopes}


\author{K. Takefuji\altaffilmark{1}, K. Sugiyama\altaffilmark{2}, Y. Yonekura\altaffilmark{3}, T. Saito\altaffilmark{4}, K. Fujisawa\altaffilmark{5} and T. Kondo\altaffilmark{1} }

\altaffiltext{1}{National Institute of Information and Communications
Technology, 893-1 Hirai, Kashima, Ibaraki 314-8501, Japan} 
\altaffiltext{2}{Mizusawa VLBI Observatory, National Astronomical Observatory of Japan, 2-21-1 Osawa, Mitaka, Tokyo 181-8588, Japan}
\altaffiltext{3}{ Center for Astronomy, Ibaraki University, 2-1-1 Bunkyo, Mito, Ibaraki 310-8512, Japan}
\altaffiltext{4}{ College of Science, Ibaraki University, 2-1-1 Bunkyo, Mito, Ibaraki 310-8512, Japan}

\altaffiltext{5}{ The Research Institute for Time Studies, Yamaguchi University, 1677-1 Yoshida, Yamaguchi, Yamaguchi 753-8511, Japan }
\email{takefuji@nict.go.jp}



\begin{abstract}
For the sake of high-sensitivity 6.7 GHz methanol maser observations, we developed a new technology for coherently combining the two signals from the Hitachi 32 m radio telescope and the Takahagi 32 m radio telescope of the Japanese Very long baseline interferometer Network (JVN), where the two telescopes were separated by about 260 m. After the two telescopes were phased as a twofold larger single telescope, the mean signal-to-noise ratio (SNR) of the 6.7 GHz methanol masers observed by the phased telescopes was improved to 1.254-fold higher than that of the single dish, through a Very Long Baseline Interferometry (VLBI) experiment on the 50 km baseline of the Kashima 34 m telescope and the 1000 km baseline of the Yamaguchi 32 m telescope.
Furthermore, we compared the SNRs of the 6.7 GHz maser spectra for two methods. One is a VLBI method and the other is the newly developed digital position switching, which is a similar technology to that used in noise-cancelling headphones. Finally, we confirmed that the mean SNR of method of the digital position switching (ON-OFF) was 1.597-fold higher than that of the VLBI method.
\end{abstract}





\section{Introduction}  
Well-known antenna arrays [e.g., Westerbork, Giant Metrewave Radio Telescope (GMRT), Australia Telescope Compact Array (ATCA), Very Large Array (VLA), and Atacama Large Millimeter/Sub-millimeter Array (ALMA)] have two observation modes:  interferometer and phased array modes. In the interferometer mode, a fine map of a radio source is obtained after the Fourier transform of the u-v plane by using  Earth's rotation. In the phased array mode, all telescopes are combined coherently into a huge single telescope not only to observe pulsars with high sensitivity (\citealt{2016MNRAS.456.2196B}) but also to join a VLBI session between distant radio telescopes.  
To combine two independent signals coherently, interferometry between two antennas of the array is first carried out with a radio continuum source such as a quasar to determine the delay among the antennas. Then, the delay, delay rate, and phase information are obtained from  all combinations of each antenna pair by the least-squares method  (\citealt{2012evn..confE..53A}) or the minimum spanning tree algorithm (\citealt{2000PASJ...52..267T}). 

Here, we focus on observing a series of maser sources with the phased-up technology. Once the delay between the two antennas is  determined by the observation of a radio continuum source, it is expected that the delay can be routinely applied to future maser observations.
However, the phase obtained by a radio continuum source will be subjected to fluctuate by the atmosphere as we observe at a different angle and  signal transfer from the antenna to the receiver and to the recorder. We determine the phase using the observed maser source itself. Since the bandwidth covering  a maser emission is generally on the order of 100 kHz, about 4.5 $\rm km\,s^{-1}$ for the velocity coverage, the accuracy of the group delay determination, which is the reciprocal of the bandwidth, will worsen.
Since it is difficult to synthesize the two signals without any corrections of the phase difference,
the establishment of a synthesis technology using two signals from two stations with a maser observation as a reference is our first goal.
 Once this technology is realized, the phase difference of the synthesis parameters can be determined by the maser observation with two antennas. As a result, the observation efficiency is expected to improve.
 
Moreover, in the case of spectral observation, the position switching  of a single dish is performed by moving the antenna physically. Then, the system noise and other background noises can be removed and only the maser signal will remain.
Once two antennas are phased, it is expected to create a virtual-off source by changing the phase of the synthesis parameters similarly to noise-cancelling headphones.
Without physically moving the antenna, the digital position switching observation is considered to be performed. This is the second goal of our research. 
Thus, we attempt to use the above techniques with the Hitachi 32 m radio telescope and  the Takahagi  32 m radio telescope of the Japanese VLBI network (JVN), which are separated by about 260 m. Furthermore, we perform VLBI experiments simultaneously with the Kashima 34 m radio telescope at a distance of about 50 km  and the Yamaguchi 32 m radio telescope at a  distance of about 1000 km  to confirm the improved sensitivity of a combined signal.
\section{Observations and data reduction}

We carried out the maser observation from 03:47 to 16:32 (UTC) on 26 Oct 2016 (DOY 300). 
Before the maser observation, we observed the quasar NRAO512 \footnote{RA=16 40 29.63277,DEC=+39 46 46.0285 (J2000) by \cite{1538-3881-150-2-58}}  to determine the geometric and clock delays among telescopes and performed a series of maser observations. 
The telescopes used in each observation are described in Table \ref{table:obs}.
The band-filtered radio frequency (RF) signals (from 6664 to 6672 MHz) of antennas were down-converted and the intermediate frequency (IF) signals (from 0 to 8 MHz) were sampled by using the analogue to digital sampler K5/VSSP32 with a 16 MHz sampling speed and 4-bit quantization (\citealt{2003ASPC..306..205K}).   

\subsection{Antennas}
The Hitachi and Takahagi radio telescopes are separated by about 260 m. The aperture efficiency of each telescope is about 55 to 70\% at 6.7 GHz depending on the elevation angle. The receivers have almost the same specifications. The two circular polarization signals from 6.3 to 8.8 GHz from the front-end receiver are divided and filtered from 6.3 to 7.0 GHz mainly for the methanol maser observation and from 8.0 to 8.8 GHz for the radio continuum observation. The two target frequency ranges can be switched within 10 min by changing the local frequency for the down-converter. A hydrogen maser was installed in Takahagi. The reference signal generated from the hydrogen maser is distributed by a coaxial cable in Takahagi and is transmitted by an optical cable to Hitachi (\citealt{2016PASJ...68...74Y}).

The Yamaguchi 32 m telescope is located in a telecommunication satellite complex of the Japanese cellular phone corporation (\citealt{2002aprm.conf....3F}). Originally, the antenna was used in the range from 4 to 6 GHz. Then, it was replaced with a newly developed broadband system from 6.3 to 8.8 GHz, which had the same specification with the Hitachi and Takahagi system. However the two frequency ranges from 6.3 to 7.0 GHz and from 8.0 to 8.8 GHz are simultaneously received with two  down-converters. The aperture efficiency is typically 65\% at 6.7 GHz.

The Kashima station is about 50 km away from  Hitachi and Takahagi, in which  receivers for the L, S/X, and K/Q bands were installed. Moreover, for the development of next-generation geodetic VLBI (VLBI Global Observing System, VGOS), a broadband system  (from 3.2 to 14.4 GHz) was installed (\citealt{RDS:RDS20528}). The linear polarization was set to be vertical in the observation. The signals from 3.2 to 14.4 GHz from the front-end receiver is transmitted by an optical cable, then the filtered signal from 6.5 to 7.0 GHz was down-converted to the IF signal. The typical aperture efficiency is about 40\% at 6.7 GHz.

\subsection{Parameters for phasing two radio telescopes by maser observation}
If we define $P_{a}$ as the signal power of antenna A and $P_{b}$ as the signal power of antenna B, then $P_{sum}$, the signal power after combining antennas A and B  with  phase difference $\Delta\theta$, can be expressed as
\begin{eqnarray}
 P_{sum} &=& \arrowvert P_{a} + P_{b} \arrowvert \\
  &=&	\sqrt{P^2_{a}+P^2_{b}+2P_{a}P_{b}\,\cos\Delta \theta \label{eqn:2} }.
\end{eqnarray}

The maximum value of the combined signal is $P_{a} + P_{b}$ and the minimum value is $P_{a} - P_{b}$ (when $P_{a} \geq P_{b}$).
If the signal powers of the two antennas are equal, the power is doubled. However,  depending on the phase difference $\Delta\theta$, the signal will decrease and disappear in the worst case. As described later, the extinction state can be used as an off-source observation. By adding an offset of $180^\circ$ to the perfectly matched angle $\Delta\theta$, it can be applied for digital position switching. 
%

By focusing on the phase, the rotation of the phase difference $\Delta\theta$, which is caused by the delay difference between the two antennas, shifts very rapidly in our experiment.  In order to achieve the phasing efficiency of 95\% (when $P_{a} = P_{b}$), the phase difference $\Delta\theta$ has to be synchronized within about 36 degree ($\sim$ 4.5 mm at 6.7 GHz) using the Eq. \ref{eqn:2}.
The predicted values such as the geometric delay and delay rate  are  calculated using a VLBI software in advance and are reasonably accurate. However, it is necessary to determine the phase difference between the antennas by actual observation.
Thus,  interferometry observations are performed by observing radio continuum sources over a  bandwidth of 10 MHz order.

On the other hand, the bandwidth covering  maser emission is generally on the order of 100 kHz. Therefore, the accuracy of the delay determined by interferometry worsens.  
In addition, if the phase at the frequency of the peak intensity of the maser emission in the cross-spectrum (in our experiment, the peak frequency can be any value from 0 to 8 MHz in the IF) is determined, it has to be converted to the phase at the sky frequency, the phase at 0 MHz in the IF.  In this conversion, the precise delay information is also required. It is difficult to synthesize the signals of two telescopes without any corrections of the phase difference.


Here, we will determine the phase difference $\Delta\theta$ by the following procedure. Firstly, we observe a strong quasar to determine the clock offset between two antennas. If the reference clocks of the telescopes are the same, the difference in the clock rate between the telescopes is  small. Otherwise, it would be better to observe the quasars in the latter scan to estimate the clock rate. Secondly, we observe a series of maser sources with the two telescopes. Thirdly, we correlate the recorded data between the two telescopes whose geometric or clock delay were adjusted to be identical so that the phase difference $\Delta\theta$ was involving the whole delay. Finally, we convert the obtained phase difference $\Delta\theta$ at the maser line to the phase at the sky frequency.  

\begin{figure}[hbtp]
  \includegraphics[width=16cm, trim= 0 0 0 0 ]{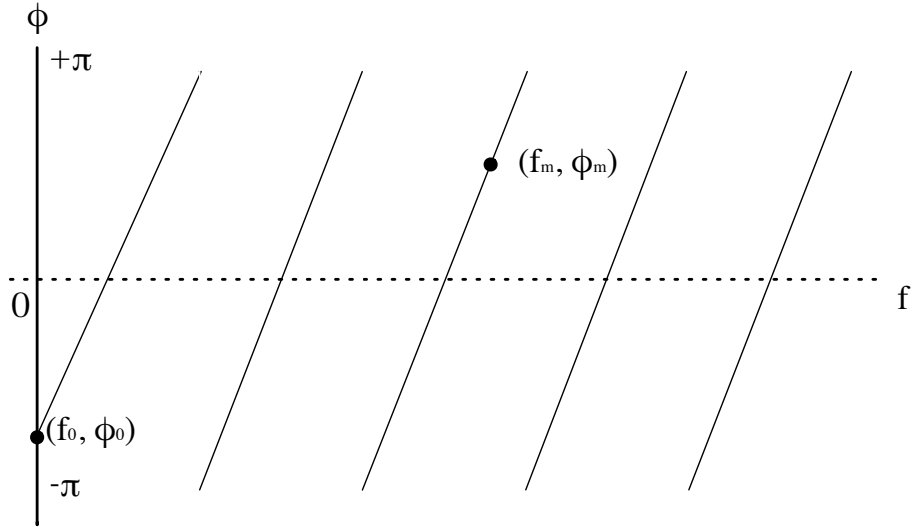}
  \caption{ Relationship of the initial phase difference $\phi_{0}$ at the sky frequency $f_{0}$ and the phase difference $\phi_{m}$ at the maser emission line $f_{m}$  against an intermediate frequency on cross-spectrum after a correlation. The discontinuous solid lines with the slope of $2\pi\tau_{g}$ represent the delay $\tau_{g}$ between
two telescopes determined beforehand by the observations of the radio continuum sources.}
  \label{fig1:phase}
\end{figure}

Figure \ref{fig1:phase} shows a conceptual phase on a cross-spectrum after performing a  cross-correlation for the phase conversion. A maser emission line is present at the frequency and the  fringe phase $(f_{m}, \phi_{m})$  at the sampling frequency of $f_{sps}$[Hz].  The initial phase  at the sky frequency $(f_{0},\phi_{0})$ is required to synthesize two telescopes. Here, the total delay comprising the geometric and clock delays is represented as the discontinuous slopes due to an ambiguity of $2\pi$  of the cross-spectrum. Now, we define the total delay as $\tau_{g}$ [s] and the initial phase is expressed as
\begin{equation}
  \phi_{0} = \phi_{m} - 2\pi f_{m} \tau_{g} \;\;(mod \;\; 2\pi). \label{eqn3}
\end{equation}
  In contrast to the continuum source observation, the phase can be determined only at the frequencies where the maser emission exist. Therefore, discontinuous slopes in Fig. \ref{fig1:phase} do not appear, so that it is difficult to convert $\phi_{m}$ into $\phi_{0}$ .
However, once the delay $\tau_{g}$ is obtained by the previous quasar observation, it becomes possible to determine $\phi_{0}$ from $\phi_{m}$ using Eq.~\ref{eqn3}. In addition, the delay rate can be determined by tracking the phase $\phi_{m}$ of the maser emission line. Thus, the parameters for phasing two telescopes can be obtained. 

\begin{figure}[hbtp]
	\centering
  \includegraphics[width=16cm, trim= 0 0 0 0]{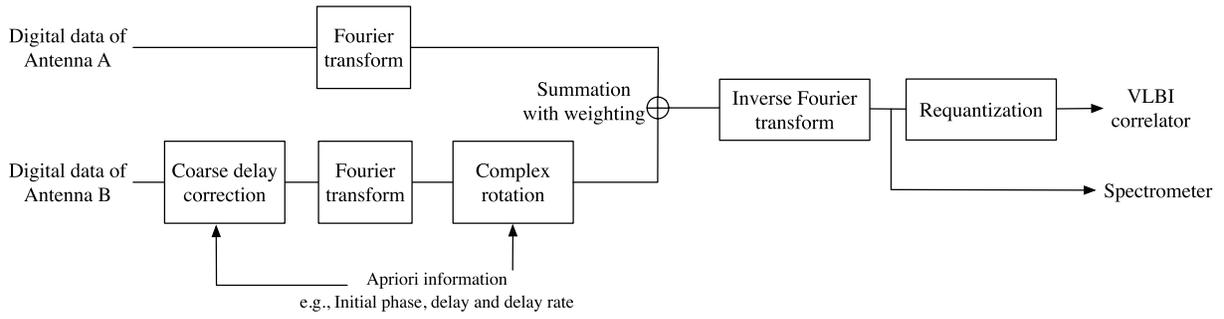}
  \caption{ Schematic diagram of the procedure of the phased telescopes.}
  \label{fig2:diagram}
\end{figure}

Figure \ref{fig2:diagram} shows a schematic diagram of the phased telescopes. Firstly, the Fourier transform is performed for two signals from the two telescopes and then the delay with reference to the antenna A is corrected in the frequency domain. We actually performed a 2048-length fast Fourier transform (FFT) for the signals of Hitachi and Takahagi. After the FFT, the obtained phase, delay, and delay rate were corrected in a frequency domain. The signal of  the antenna B at this time can be considered to have been observed at the position of the antenna A. Thereafter, signals were synthesized and returned to the time-series data by inverse FFT (2048 points). Finally, the combined signal was generated. 
\section{Results }
\subsection{Phased telescopes}
\begin{figure}[hbt]
\centering
  \includegraphics[width=16cm, trim= 0 0 0 0]{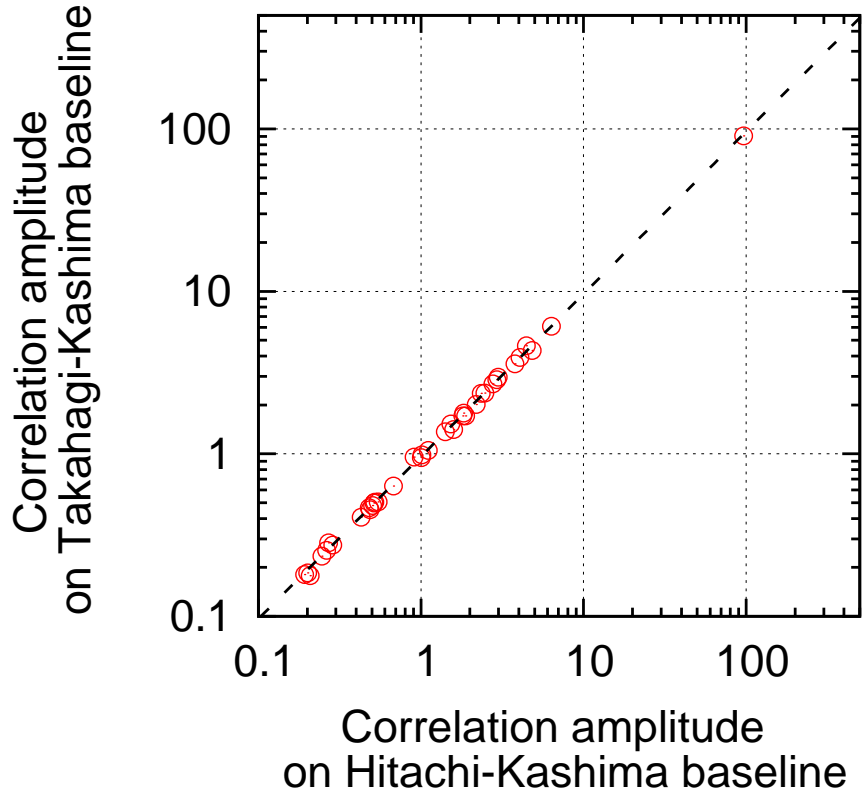}
  \caption{Comparison of the correlation amplitudes of the maser sources of the Takahagi and Kashima baseline against  the Hitachi and Kashima baseline. The mean correlation amplitude of the Takahagi baseline relative to that of the Hitachi baseline was 0.941$\pm$0.102 (dashed line).}
  \label{fig3:compare-amplitude}
\end{figure}

\begin{figure}[hbtp]
  \includegraphics[width=16cm, trim= 0 0 0 0]{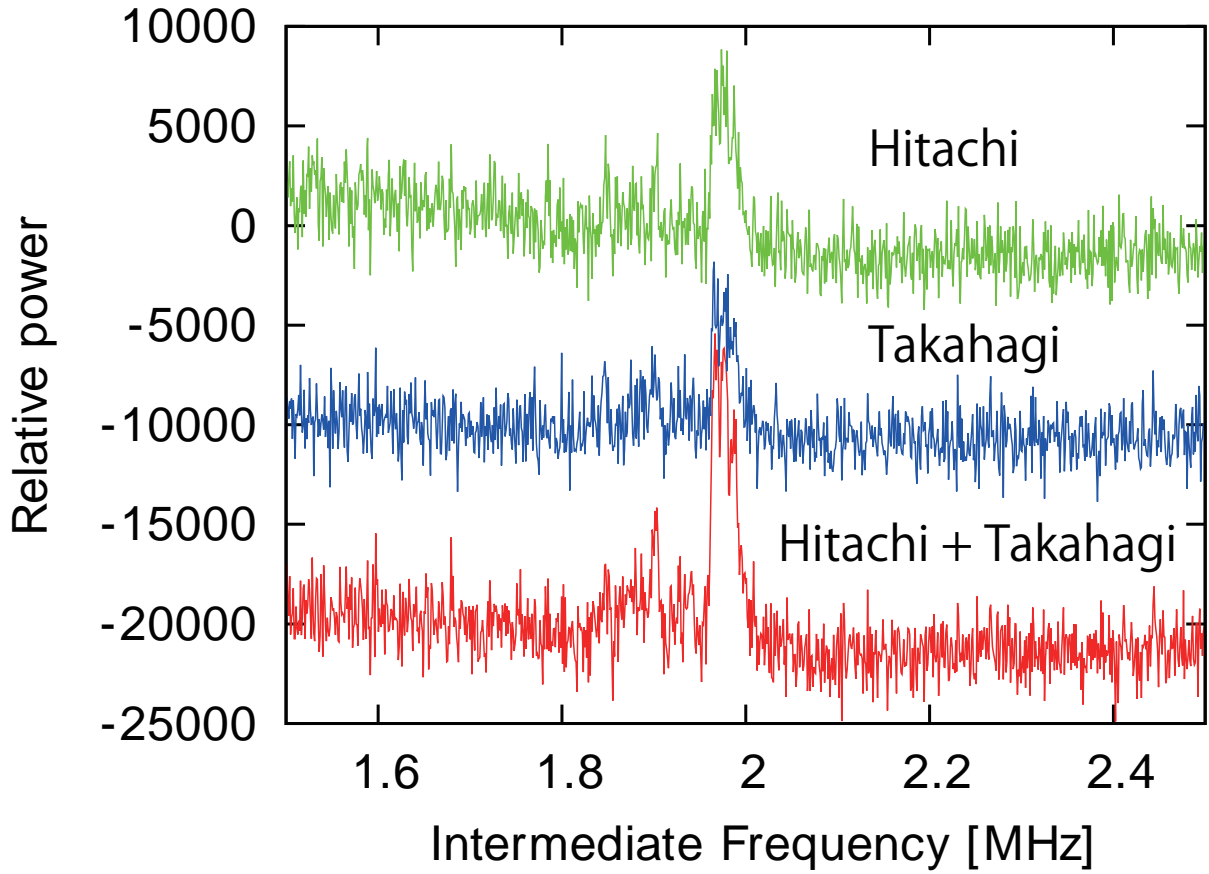}
  \caption{Spectra of the maser source G26.52-0.26 obtained by the Hitachi telescope (top),  by the Takahagi telescope (middle) and by the phased Hitachi and Takahagi telescopes (bottom) with 60 s integration and 1 kHz resolution. The SNRs of each figure were estimated to be about 4.8, 4.6 and 6.4, respectively. }
  \label{fig10:phased}
\end{figure}

Before phasing of the Hitachi and Takahagi antennas, we performed a common correlation to evaluate their correlation amplitudes with the software correlator GICO3.  
Figure \ref{fig3:compare-amplitude} shows the correlation amplitude on the baseline of Hitachi and Kashima against those on the baseline of Takahagi and Kashima, where only detected masers were plotted. The sensitivities of the Hitachi and Takahagi antennas were similar throughout the observation. The mean correlation amplitude of the Takahagi baseline relative to that of the Hitachi baseline was 0.941$\pm$0.102. If the sensitivities of the two telescopes are different, the signal needs to be weighted during the synthesizing process.
In accordance with the aforementioned procedure, we observed a total of 56 masers using the Hitachi and Takahagi telescopes as the phased antennas. The recorded data of the Takahagi 32m telescope were shifted to those of the Hitachi 32m telescope position. Figure \ref{fig10:phased} shows the spectra of the maser source G26.52-0.26 with the phased  and unphased antennas with 60 s integration and 1 kHz resolution. We removed the trends of the bandpass profiles by the least-squares method with a quadratic function and give the offset for comparison. With regard to the SNR calculation, the noise was determined  by the standard deviation of the band profile in the IF range of  2.05 to 2.2 MHz. By adding  6664 MHz, the IF signal can be converted to the RF signal. The SNRs of each spectrum were estimated to be about 6.4, 4.8, and 4.6 from top to bottom.
%
%
\subsection{Digital position switching} \label{ch:3}
When we take the difference between the on-source and the off-source in a single-dish observation, we physically move the antenna from the direction of the radio source to a blank sky in a normal manner. By taking the difference between the on-source and the off-source, 
the common band-pass profile and other common signals are theoretically removed and only the signal of the radio source remains. Therefore, the observation efficiency is reduced by about half owing to the time for the off-source observations.  By adding an offset of $180^\circ$ to the perfectly matched phase of the phased antenna, it can be considered as a virtual off-source and applied for digital position switching. If digital position switching is realized, the off-source observations can be skipped and therefore it is expected that the observation efficiency will be increased. 

\begin{figure}[hbt]
  \includegraphics[width=16cm, trim= 0 0 0 0]{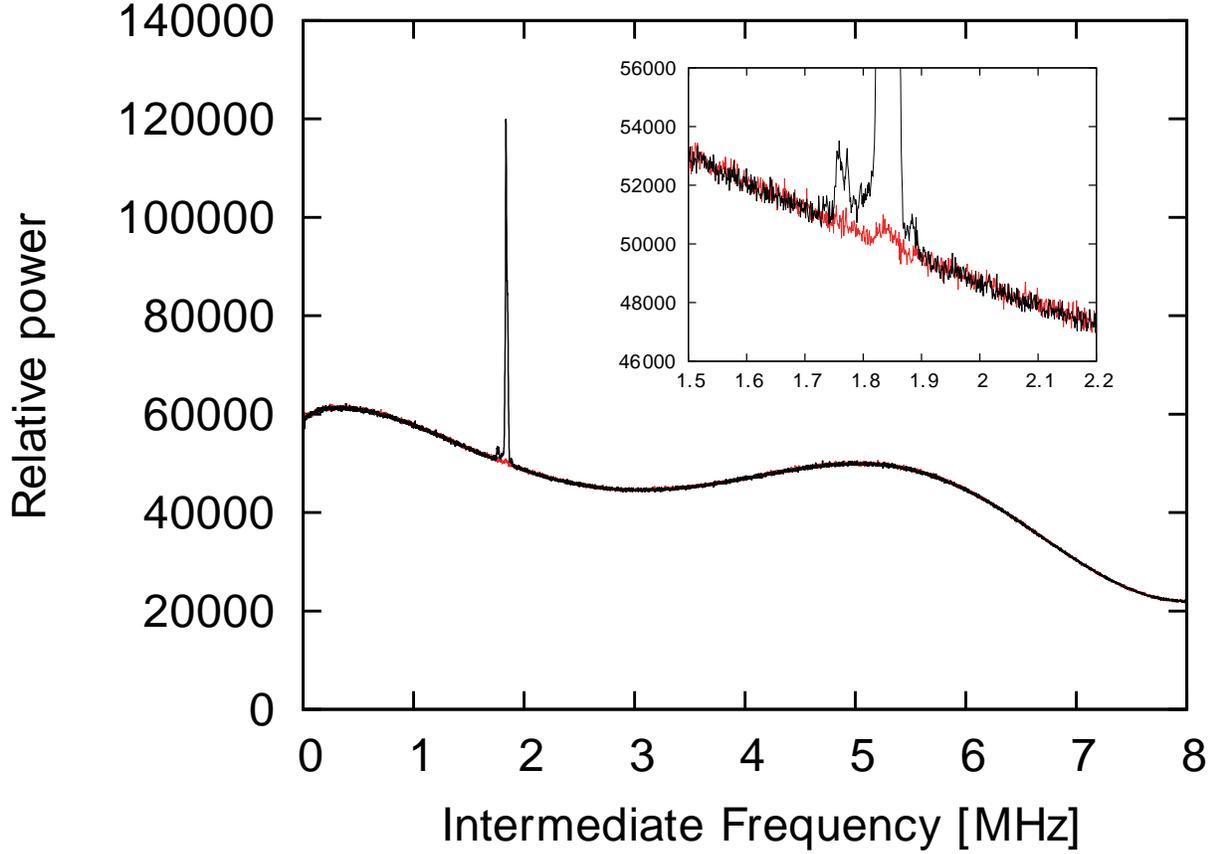}
  \caption{The on-source and the off-source spectra of maser G12.025-0.031 calculated by the digital position switching of the phased Hitachi and Takahagi antennas with 60 s integration and 1 kHz resolution. The black plot shows the on-source spectrum for in-phase combination of the two radio telescopes. The red plot shows that of the off-source obtained by suppressing the maser signal by applying a phase difference of $180^\circ$ to the in-phase combination. Plot in the upper-right is the close-up baseline around the frequency of the maser emission.}
  \label{fig4:in-phase}
\end{figure}

Figure \ref{fig4:in-phase} shows the on-source and off-source spectra calculated using the phased  Hitachi and Takahagi antennas. The on-source data contains the maser signal of G12.025-0.031 for the in-phase combination with 60 s integration and 1 kHz resolution. On the other hand, the off-source data is obtained by suppressing the maser signal by applying a phase difference of $180^\circ$ to the in-phase combination. From the figure, the band characters without the maser region have the same curvature and were consistent between the on-source and the off-source. In addition, the weak maser signal, which was observed in an enlarged view of the off-source,  had about 0.8\% of the amplitude of the on-source. This might be caused by a fluctuation of the signal transfer and a change in the atmosphere thickness, which causes a small phase difference between the two stations. Here, we provided a single parameter for the phasing procedure. However, we think that the amplitude of the off-source can be reduced further by providing synthesis parameters over a short time. Moreover, the amplitude of the on-source can be increased owing to the more accurate accumulation.

\begin{figure}[hbt]
  \includegraphics[width=16cm, trim= 0 0 0 0]{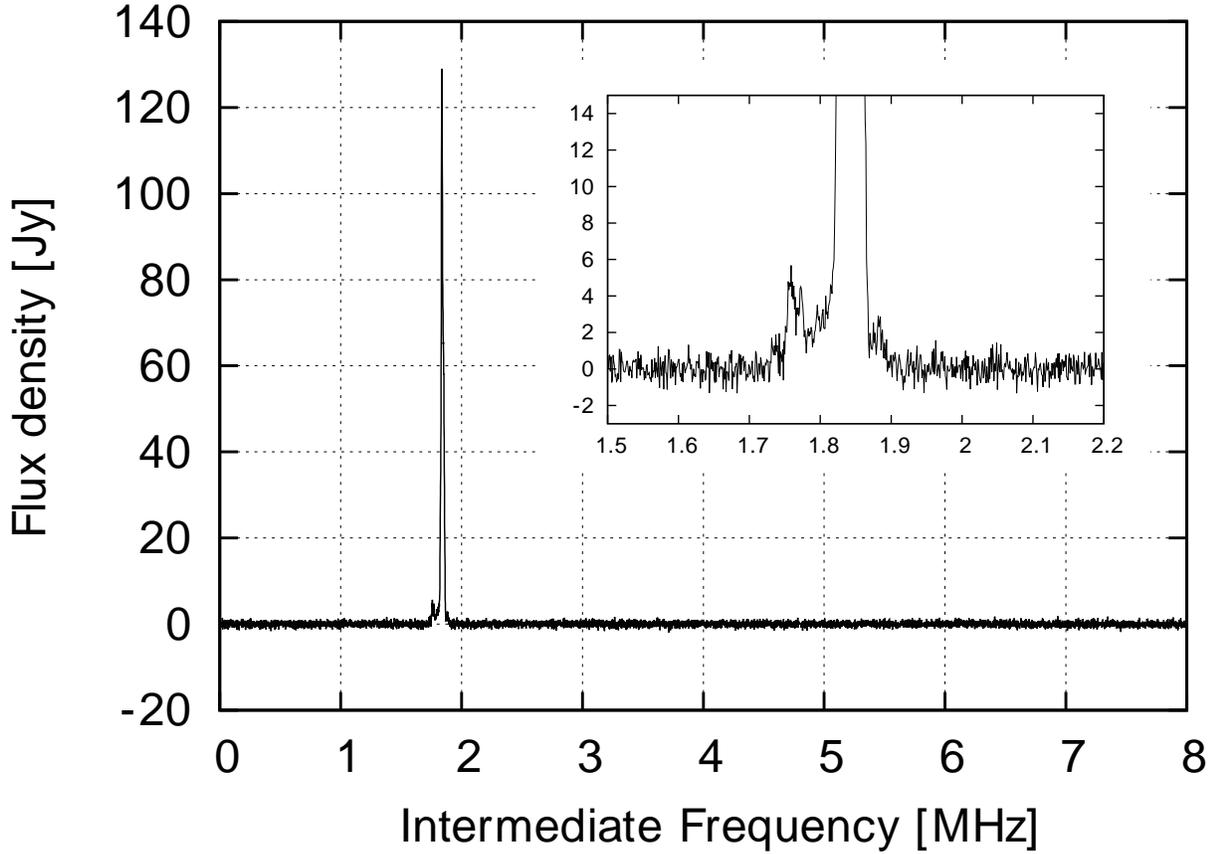}
  \caption{Spectrum formed from the ratio (ON-OFF)/OFF, where the ON and OFF data are total power samples from figure \ref{fig4:in-phase}. We finally obtained the flux density of the maser source  G12.025-0.031 with the SEFD assumed to be 85 Jy for converting from the relative amplitude to the flux density.  The background noises (e.g., band character due to the receiver system) were clearly removed. Plot in the upper-right is the close-up baseline around the frequency of the maser emission.}
  \label{fig5:monopulse} 
\end{figure}

Figure \ref{fig5:monopulse} is obtained by multiplying the ratio (ON-OFF)/OFF by the system equivalent flux density (SEFD), which we assumed to be 85 Jy. Since the SEFD of both the Hitachi and Takahagi telescopes is approximately 170 Jy, when the efficiency and the system temperature are 70\% and 30 K, respectively,  we provided the half value of the SEFD. The system temperature has 10 to 20\% measurement error, so that an error propagates in the calculation of the flux density. Here, the obtained flux density of the maser sources is consistent  with our recent monitoring result. Since the power, when the maser did not exist, is close to zero, the common components such as the band character were removed. However, if a maser source is in the strong HII region,  the contribution  of the continuum emission such as that in the HII region will remain even after the on-off processing. 

\begin{figure}[hbtp]
  \includegraphics[width=16cm, trim= 0 0 0 0]{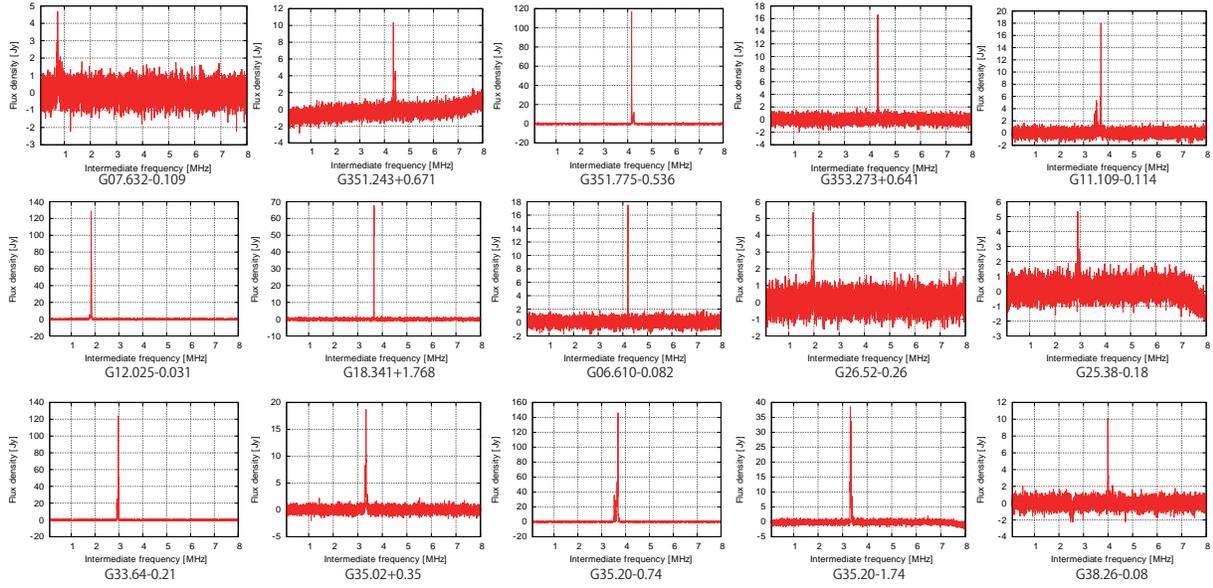}
  \caption{ Sample spectra of maser sources obtained by the phased telescopes of Hitachi 32 m and Takahagi 32 m after digital position switching was performed with 60 s integration and 1 kHz resolution. }
  \label{fig11:maser1}
\end{figure}

Figure \ref{fig11:maser1} shows the maser spectra of the phased Hitachi and Takahagi telescopes after the digital position switching was performed with 60 s integration and 1 kHz resolution.  Details of the maser observations will be presented in separate papers. However, 41 maser sources from 56 maser observations were detected by the digital position switching process. We confirmed that the phased telescopes operated over 12 hours with the maser observation as the reference. On the other hand, the residual powers of four maser sources (G351.243+0.671, G25.38-0.18, G35.20-1.74, and G43.87-0.77) were not close to zero. 
When a signal from an extended centimeter continuum source (e.g., an HII region)  is contaminated with a maser signal, the spatial structure of the continuum source produces a phase residual to both the on-source and off-source. Thus, a power difference should appear. Actually the correlation of the four maser observations suggested the existence of a continuum emission. Previous researches (G351.243+0.671 by \citealt{2008MNRAS.386.1521C}, G25.38-0.18 and G43.87-0.77 by \citealt{2016ApJ...833...18H}, and G35.20-1.74 by \citealt{1994ApJ...426..249O}) reported the continuum emissions toward the four maser sources, which have a spatial structure almost equal to the spatial resolution of the phased Hitachi and Takahagi telescopes ($\sim$ 35 arcsec).  If we assume that the continuum emission is additively composes to the on-source or the off-source, we apply the following ratio of the on/off to make  the residual power zero: 102\% for G351.2+0, 97\% for G25.38-0.18, 97\% for G35.20-1.74, and 99\% for G43.87-0.77.           
	
\section{Discussion}

\subsection{Sensitivity comparison of the maser signals of a single antenna with phased antennas}
\begin{figure}[hbt]
  \includegraphics[width=16cm, trim= 0 0 0 0]{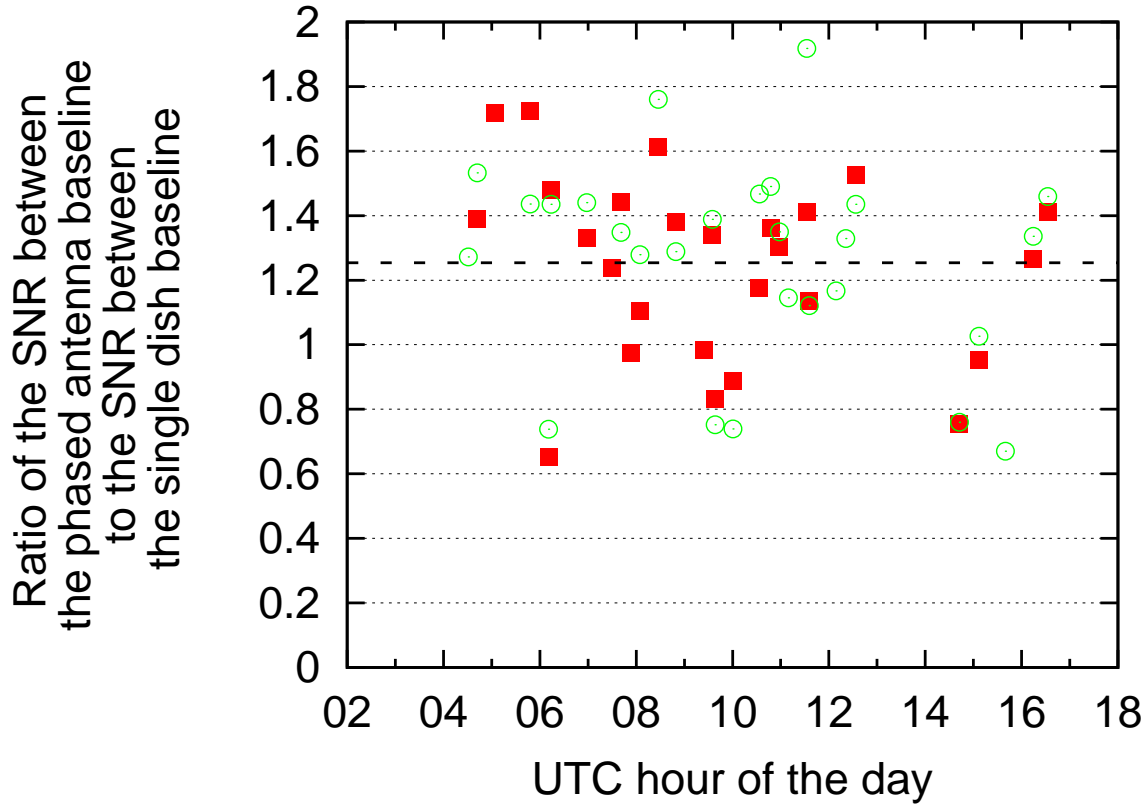}
  \caption{Comparison of the maser signal-to-noise ratio (SNR) of VLBI of the single dish and the phased two telescopes against Kashima 34 m and Yamaguchi 32 m as a function of time. 
     The filled squares show the maser SNR ratio of VLBI against the Kashima 34 m radio telescope and the opened circles show that against the Yamaguchi 32 m radio telescope. The mean SNR ratio is 1.254 $\pm$ 0.298 (dashed line).}
  \label{fig6:vlbi} 
\end{figure}

The phased antenna, which is obtained by  coherently combining by two antennas of similar sensitivity, is expected to have a twofold larger aperture. Then,   
if we carry out VLBI between the phased antenna and another radio telescope, the SNR will become $\sqrt{2}$ times better than that of the single antenna because the SNR of VLBI is proportional to the square root of the product of the apertures of the two antennas.
Here, we perform cross-correlation processing of the observed maser sources. The result of the sensitivity comparison is shown in Fig. \ref{fig6:vlbi}. We set identical correlation parameters (e.g.,  the clock offset, the geometric delay, the delay rate, and the frequency resolution of 1 kHz) and extracted the results with an SNR of over 15. Since the noise of the power in the Fourier domain follows the Rayleigh distribution, the expectation of the noise, when the variance of both the real and imaginary part is $\sigma$, becomes $\sigma\sqrt{\pi/2}$ ($\sim 1.25\sigma$). Thus, it was required to give the larger thresholds of the SNR than the usual Gaussian data and we empirically set the thresholds as 15. 
From the results, the mean SNR of the phased antenna becomes 1.254-fold improved. As mentioned above, we expected that the improvement would be $\sqrt{2}$ $(\sim 1.414)$. We assume that this difference was caused by the requantization after synthesizing the two recorded data. With regard to the 2-bit quantization that we applied, a factor of 0.881 is multiplied to the SNR \citep{Thompson2017}. Therefore, we confirmed that the phased antenna has about a twofold larger aperture.

\subsection{Sensitivity comparison of the maser signals of VLBI  with  phased antennas}
According to \cite{gmrt}, if we observe a radio source of  flux density $S$ in a integration time $t$ with a single dish whose gain, bandwidth, and system temperature are $G$, $B$, and $T_{sys}$, respectively, the SNR can be expressed as
\begin{eqnarray}
	SNR_{single} = \frac{ GS\sqrt{Bt}}{T_{sys}}. 
\end{eqnarray}
 In VLBI, an interferometer is composed of two antennas $i$ and $j$. Then, the SNR becomes \citep{Thompson2017,Takahashi200001,  gmrt} 
  \begin{eqnarray}
	SNR_{VLBI} = \frac{ GS\sqrt{2Bt}}{\sqrt{T_{sys_{i}}T_{sys_{j}}}}.
\end{eqnarray}    
 
\begin{figure}[hbtp]
  \includegraphics[width=16cm, trim= 0 0 0 0]{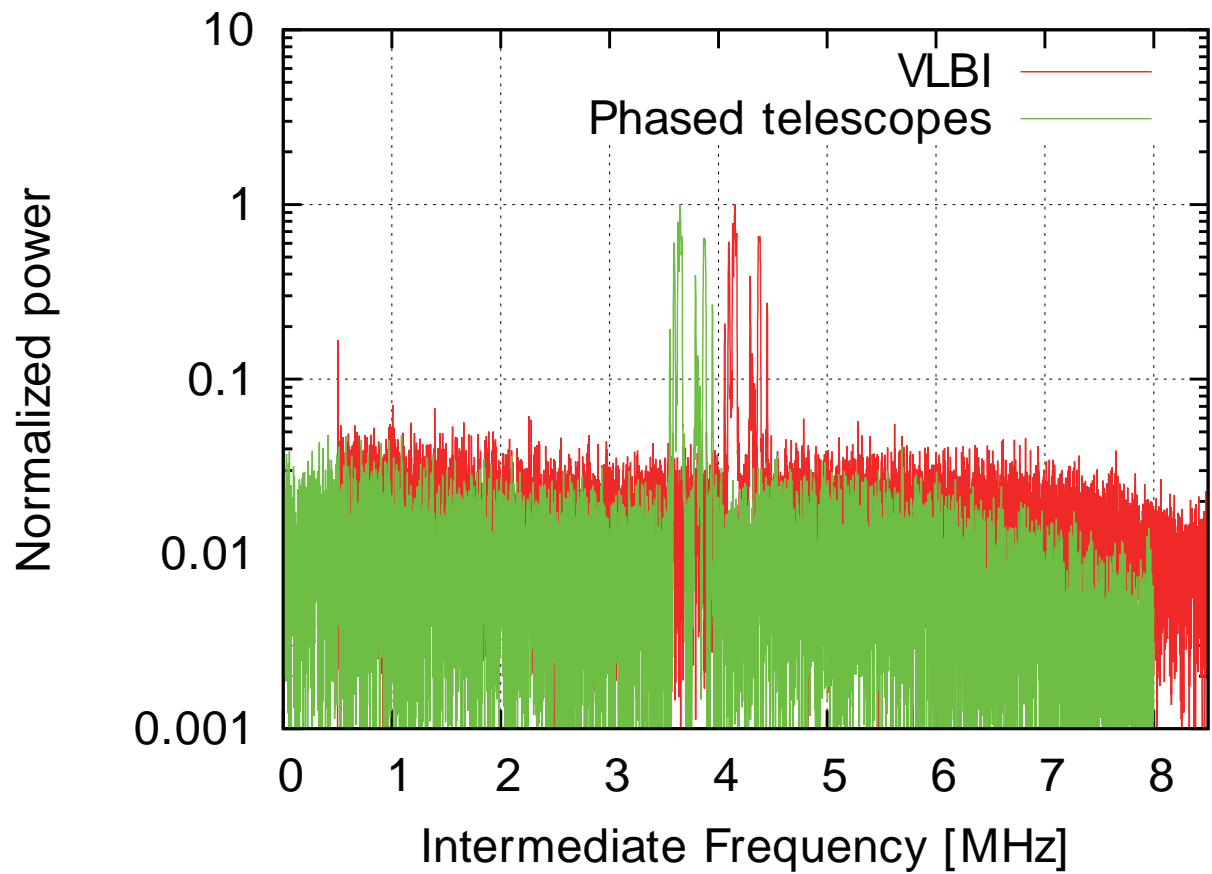}
  \caption{ Comparison of the spectra of the maser source G06.795-0257 between VLBI (red) and the phased telescopes (green) of Hitachi 32 m and Takahagi 32 m. The frequency resolution and integration time were unified to  1 kHz and 60 s, respectively.   The plots were normalized by the peak power of the maser. The spectrum of the phased telescopes have a lower noise level than that of VLBI across the whole bandwidth. The SNR of VLBI is 104, compared with 136 for the phased telescope mode. The plot of VLBI was +0.5 MHz shifted for clarity.}
  \label{fig9:single}
\end{figure}

 For a single dish with the collecting area equal to the sum of two identical antennas $i$ and $j$, with  twice as large gain $2G$ and system temperature $T_{sys}=\sqrt{T_{sys_{i}}T_{sys_{j}}}$, the SNR should be a factor of $\sqrt{2}$ better  than that of VLBI \citep{gmrt}. Thus, once two telescopes are phased, the sensitivity should be better than that of VLBI. Figure \ref{fig9:single} shows a comparison of the spectra of the maser source G06.795-0.26 between the VLBI (cross-spectrum) and the phased telescopes of Hitachi 32 m and Takahagi 32 m. The frequency resolution and integration time were unified to be 1 kHz and 60 s, respectively, for careful comparison. For the case of VLBI, data of Hitachi and Takahagi were correlated  in the usual manner, while in the phased method, the spectra of ON and OFF were obtained by digital position switching  and then the power of ON-OFF were plotted. In section \ref{ch:3}, the spectra of the detected masers were processed by calculating (ON-OFF)/OFF. However, the division by the off-source makes the band profile flat. Hence, we skipped the division by the off-source for careful comparison.  Finally, both data were normalized by the peak power of the maser emission. The spectrum of the phased telescopes seems to have a lower noise level  than that of VLBI across the whole bandwidth, and the SNR of VLBI is 104, compared with 136 for the phased telescope.

\begin{figure}[hbtp]
  \includegraphics[width=16cm, trim= 0 0 0 0]{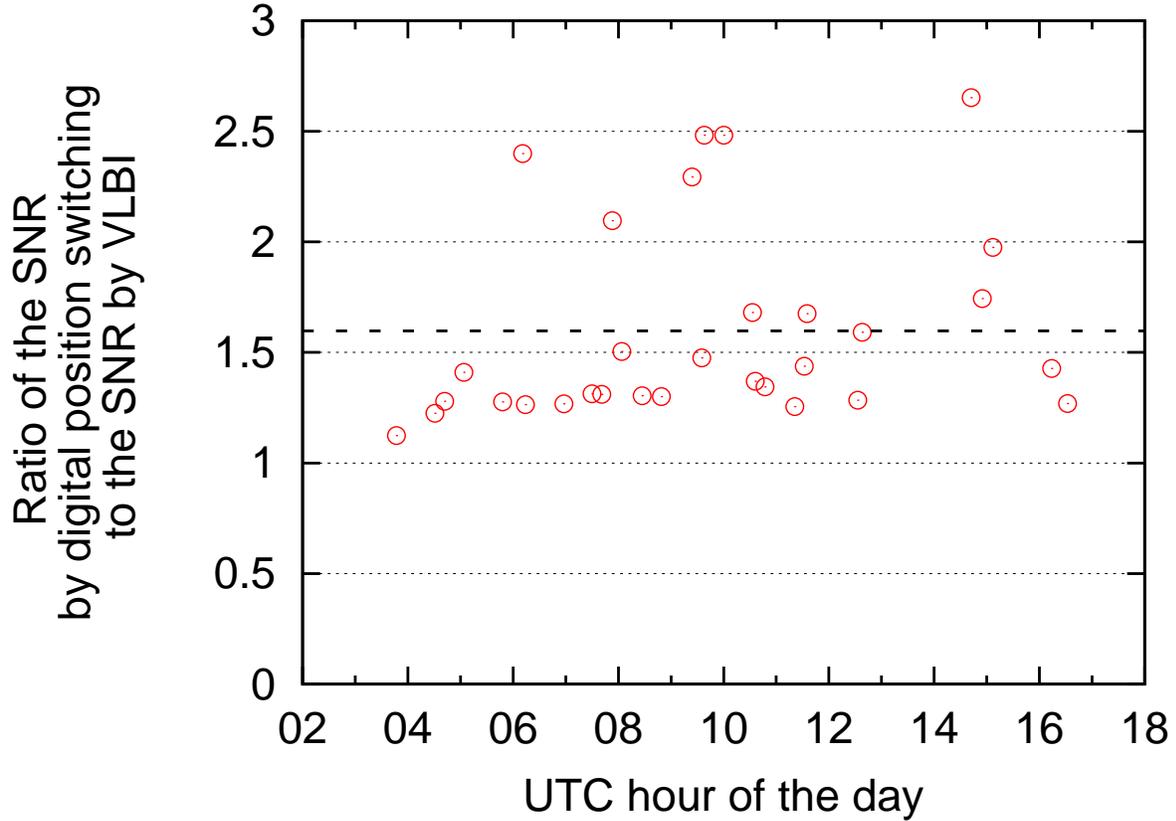}
  \caption{Comparison of the maser SNR of VLBI and the phased telescopes of Hitachi 32 m and Takahagi 32 m as a function of  time. The SNRs by the VLBI method were extracted from the cross-spectrum of the VLBI made for the Hitachi and Takahagi telescopes. While, the SNRs by the phased method were obtained by the spectrum of digital position switching (ON-OFF) with the phased two telescopes of Hitachi and Takahagi telescopes. The mean SNR ratio of the phased antenna relative to that of VLBI is 1.597 $\pm$0.444 (dashed line) and the median of the SNR ratio is 1.410.}
  \label{fig16:vlbi-onoff}
\end{figure}

 Finally, Fig. \ref{fig16:vlbi-onoff} shows a comparison of the SNRs of the detected maser sources with 60 s integration obtained using VLBI and the phased telescope. Only the data whose SNR was over 15 were plotted. The mean SNR of the phased antenna relative to that of VLBI is 1.597$\pm$0.444. In the calculation of the SNR, the $1\sigma$ noise was commonly defined for both methods from the scan, in the frequency range where no maser lines were detected. Then, the SNR was obtained by dividing the peak power of the maser line by the defined $\sigma$. The spectra obtained by the digital position switching were produced without the requantization effect. Thus, the obtained SNRs could be directly compared. Since some maser scans had a factor of 2 or 3 unexpectedly, the results may have been overestimated (the median of the SNR ratio was 1.410). We assume that  the radio continuum emission toward the maser source is one possibility.  Consequently, we confirmed the effectiveness of the digital position switching. Furthermore, we plan not only to synthesize three or more telescopes but also to evaluate other frequency bands (e.g., a water maser at 22 GHz) in future.
  
\section{Summary}   
 We developed phased-up technology using the closely positioned two telescopes Hitachi and Takahagi and applied it to a series of 6.7 GHz methanol maser observations. We confirmed that the mean SNR was improved to  1.254-fold higher than that of a single dish through a VLBI experiment on the 50 km baseline of the Kashima 34 m telescope and the 1000 km baseline of the Yamaguchi 32 m telescope.  
Furthermore, with the newly developed digital position switching, the sensitivity of  the received maser clearly increased and  the mean SNR of the digital position switching was 1.597-fold higher than that obtained by VLBI. Moreover, once the clock offset was determined by a quasar observation, we confirmed that the phased telescopes operated over 12 hours with a maser observation as a reference without any off-source observations. Thus, the off-source observations can be skipped and therefore we expect that the observation efficiency will be increased.
 
\bibliography{ms} 
\bibliographystyle{chicago}

\begin{table}[htbp]
\begin{center}
\begin{tabular}{c|cccc} 
Station & Hitachi & Takahagi & Kashima & Yamaguchi \\\hline
Diameter [m] & 32 & 32 & 34 & 32 \\
Tsys*[K] & 29.5(@EL=85 deg) & Typ 30-40  & Typ 150  & 110(@EL=46 deg) \\
Efficiency [\%] & 55-75 & 55-75 & Typ 40 &  Typ 65\\
Polarization & LHCP & LHCP & Vertical & LHCP \\
Weather& Fine & Fine & Fine & Cloudy \\
\end{tabular}
\end{center}
\caption{Parameters of the 6.7 GHz maser experiment ($fc = 6668$ MHz) on 26 Oct 2016 (DOY 300). }
\label{table:obs}
\end{table}
\clearpage

\clearpage
\clearpage
\clearpage
\clearpage

\clearpage
\clearpage

\end{document}